# A systematic review of fuzzing based on machine learning techniques

Yan Wang[a], Peng Jia[a], Luping Liu[b], Jiayong Liu[a][1]

[a] *College of Cybersecurity Sichuan University*，*No.24 South Section 1, Yihuan Road, Chengdu, China*

[b] *College of Electronics and Information Engineering Sichuan University*，*No.24 South Section 1, Yihuan Road, Chengdu, China*

**Abstract-** Security vulnerabilities play a vital role in network security system. Fuzzing technology is widely used as a vulnerability discovery technology to reduce damage in advance. However, traditional fuzzing techniques have many challenges, such as how to mutate input seed files, how to increase code coverage, and how to effectively bypass verification. Machine learning technology has been introduced as a new method into fuzzing test to alleviate these challenges. This paper reviews the research progress of using machine learning technology for fuzzing test in recent years, analyzes how machine learning improve the fuzz process and results, and sheds light on future work in fuzzing. Firstly, this paper discusses the reasons why machine learning techniques can be used for fuzzing scenarios and identifies six different stages in which machine learning have been used. Then this paper systematically study the machine learning based fuzzing models from selection of machine learning algorithm, pre-processing methods, datasets, evaluation metrics, and hyperparameters setting. Next, this paper assesses the performance of the machine learning models based on the frequently used evaluation metrics. The results of the evaluation prove that machine learning technology has an acceptable capability of categorize predictive for fuzzing. Finally, the comparison on capability of discovering vulnerabilities between traditional fuzzing tools and machine learning based fuzzing tools is analyzed. The results depict that the introduction of machine learning technology can improve the performance of fuzzing. However, there are still some limitations, such as unbalanced training samples and difficult to extract the characteristics related to vulnerabilities.

## 1. Introduction

Vulnerabilities often refer to the flows or weaknesses in hardware, software, protocol implementations, or system security policies that allow an attacker to access or compromise the system without authorization, and have become the root cause of the threats toward network security. WannaCry ransomware attack outbroke on May 2017, and more than 150 countries and 300,000 users were attacked, causing more than $8 billion in damage (Wikipedia 2019b). The virus spread widely by utilizing the "Eternal Blue" vulnerability of the NSA (Nation Security Agency) leak. The number of vulnerabilities announced by CVE (Common Vulnerabilities & Exposures) began to explode in 2017, from the original highest 7946 vulnerabilities in 2014 to the publication of 16555 vulnerabilities in 2018(CVE 2019).

Considering the increasing number and the severe damages of vulnerabilities, vulnerability discovery technology has attracted widespread attention. Fuzzing technology is an efficient method to discover weaknesses, which was first proposed by Miller et al. in 1990(Miller et al. 1990). It is an automatic testing technique that covers numerous boundary cases using invalid data (e.g., files, network packets, program codes) as application input to ensure the absence of exploitable vulnerabilities (Oehlert 2005). Since then, a variety of different techniques were proposed to improve the efficient of fuzzing. These techniques include static analysis (Sparks et al. 2007; Kinder et al. 2008), dynamic analysis (Höschele and Zeller 2016; Bastani et al. 2017; Kifetew et al. 2017), However, fuzzing test still faces

---

E-mail addresses:

[1] corresponding author: ljy@scu.edu.cn (J. Liu),
yanwang@stu.scu.edu.cn (Y. Wang)
uoscujp@gmail.com (P. Jia),
lupingllp@gmail.com (L. Liu)

many challenges, such as how to mutate seed inputs, how to increase code coverage, and how to effectively bypassing verification (Li et al. 2018).

With the advancement of machine learning in the field of cybersecurity, it has also been adopted by many studies for vulnerability detection. (Grieco et al. 2016; Wu et al. 2017; Chernis and Verma 2018), including the applications in fuzzing (Godefroid et al. 2017; Rajpal et al. 2017; Wang et al. 2017; She et al. 2018; Liu Xiao, Prajapati, Rupesh, Li Xiaoting 2019). Machine learning technology is introduced into fuzzing to provide a new idea for solving the bottleneck problems of the traditional fuzzing technology and also makes the fuzzing technology intelligent. Using machine learning for fuzzing testing will become one of the critical points in the development of vulnerability detection technology with the explosive growth of machine learning research.

However, there is no systematic review of machine learning based fuzzing in the past few years. We argue that it is necessary to write a comprehensive review to summarize the latest methods and new research results in this area. The paper aims to discuss, analyze, and summarize the following problem:

- RQ1: Why machine learning techniques can be used for fuzzing?
- RQ2: Which steps in the fuzzing have used machine learning techniques?
- RQ3: Which machine learning algorithms have been used for fuzzing?
- RQ4: Which techniques are used for data pre-processing of fuzzing based on machine learning?
- RQ5: Which datasets are used for testing and evaluating?
- RQ6: Which performance measures are used for evaluating the results?
- RQ7: How to set the hyperparameters of the machine learning models?
- RQ8: What are the performances of the machine learning models?
- RQ9: What is the capacity of discovery vulnerabilities of the fuzzing model based on machine learning technology?

The rest of the paper is organized as follows: Section 2 describes the general process and limitations of traditional fuzzing. Section 3 introduces the machine learning and summarizes the factors that lead the machine learning technology can be introduced to fuzzing test. Section 4 analyzes and summarizes the different scenarios of machine learning for fuzzing test by discussing RQ2. Section 5 systematically studies the machine learning models used in fuzzing by discussing RQ3, RQ4, RQ5, RQ6, and RQ7. Section 6 assesses the performance of machine learning-based fuzzing technology by discussing RQ8 and RQ9. Section 7 provides conclusions and future directions obtained from this systematic review.

## 2. Traditional Fuzzing Techniques

### 2.1. Working Process of Fuzzing

The processes of traditional fuzzing test are depicted in Fig. 1. Some of the steps (e.g., Testcase Filter) in fuzzing workflow may not be available in some fuzzing tools, but the general steps are the same. The working process of fuzzing is composed of four main stages, the testcase generation stage, program execution stage, runtime status monitoring, and analysis of crashes.

The testcase generation stage includes seed file generation, mutation, testcase generation, and testcase filter. The seed file is the original sample that conforms to the input format of the program. A large number of testcases are generated by selecting different mutation strategies to mutate the seed files at different locations (Lcamtuf 2014). However, we need to select testcases that can trigger new paths or vulnerabilities through the testcase filter due to not all cases are valid. The selection process is guided by the defined fitness functions, such as code coverage. The testcase generation stage can be divided into mutation-based and generation-based generation. The mutation-based generation strategy refers to the generation of new testcases based on the modification of known testcase, and it contains the entire four processes above. The generation-based generation strategy is to generate new test inputs directly based on the known input case format, and it does not contain the process of mutation above.

The program execution stage is mainly to input the generated test cases into the target program for execution. A target program is a program under test (PUT). It can be a binary program with or without a source program.

The runtime state monitoring stage monitors the state of the program at runtime and guides the generation of the test samples by feeding back the information at the time of execution to the testcase generation phase (Serebryany and Bruening 2012; Slowinska et al. 2012). The techniques used in the monitor include binary code instrumentation (Luk et al. 2005), taint analysis (Drewry and Ormandy 2007), etc. When a target program crashes or reports some error, the related information must be collected for later replay and analysis.

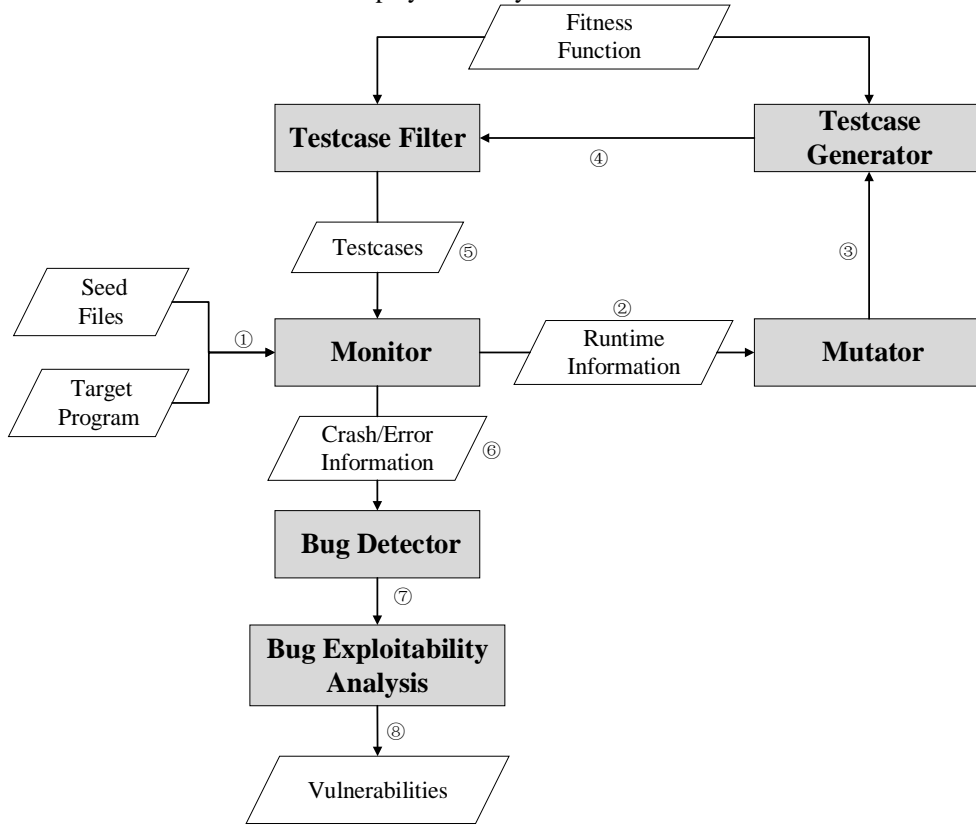

**Fig. 1. Working process of fuzzing**

In the analysis stage, the information collected at the time of the crash will be analyzed to determine whether or not the crash is a bug. Then, fuzzer classify bugs by exploitability analysis to determine whether it is a vulnerability (Chen et al. 2013; Team MSECMSS 2013; Zalewski M 2016). Finally, analysts make the final confirmation through debugging.

**2.2. The Limitations of Fuzzing**

The traditional fuzzer has three key challenges: (1) how to mutate seed inputs, (2) how to improve code coverage, (3) how to bypass the validation (Li et al. 2018). Various auxiliary analysis techniques were introduced to fuzzing test to alleviate these challenges and improve the efficiency of vulnerability detection. The introduction of these technologies has promoted the development of fuzzing technology and made the traditional fuzzing process more intelligent. However, these technologies also have certain limitations and disadvantages.

Static analysis extracts relevant information from a binary code or source code to direct the fuzzing process without running the program (Wichmann et al. 1995). However, static analysis lacks context information during the running and has a high dependence on prior knowledge, resulting in low accuracy and a high false position rate.

Dynamic analysis needs to execute the target program in real systems or emulator (Ernst 2003). It monitors program status and obtains relevant runtime knowledge at execution time. It includes dynamic symbolic execution (Cadar et al. 2006; Godefroid et al. 2008; Chipounov et al. 2011) and dynamic taint analysis (Drewry and Ormandy 2007; Haller et al. 2013; Neugschwandtner et al. 2015).

Dynamic symbolic execution generates a set of path constraints by using symbolic values as input during program execution (King 1976). Then determine whether the path is reachable and generate the corresponding test input by satisfying model theory (SMT) solver. While there exist many limitations on symbolic execution, such as path explosion (Xie et al. 2009), environment interactions (Cadar et al. 2008; Avgerinos et al. 2014), memory modeling (Cha et al. 2012; Shoshitaishvili et al. 2015), and parallel computing (Bucur et al. 2011; Avgerinos et al. 2016).

Dynamic taint analysis uses the tagged data as input to the program, which records how the program uses the input data and which program elements were tainted by the input data. However, taint analysis has the problems of under-tainting and over-tainting (Kang et al. 2011).

Considering the challenges of the state-of-the-art fuzzer and the shortcomings the various techniques already had, fuzzing testing requires a combination of new technologies and methods as countermeasures to these challenges.

## 3. Introduction of Machine Learning Technology

Machine learning acquires new knowledge or skills by learning from existing example data or experiences and automatically optimize the performance of the computer system itself (Witten et al. 2016). Machine learning tasks can be categorized into traditional machine learning, deep learning, and reinforcement learning. Traditional machine learning is divided into supervised learning, unsupervised learning, and semi-supervised learning according to whether the input data is labeled or not and the amount of labeled data.

Deep learning (Deng et al. 2014; LeCun et al. 2015) is an artificial neural network composed of multiple nonlinear processing units to representation learning of data, which is a deeper extension of the machine learning algorithm. The original "feature engineering" is replaced based on the method of representational learning, and the machine can automatically extract useful features from the input data. Reinforcement learning (Sutton et al. 1998; Sutton and Barto 2018) is a branch of machine learning methods, which describes and solves the problem of agents maximizing feedbacks or achieving specific goals through learning strategies in the interaction with the environment. Unlike learning from a large number of input samples, reinforcement learning is essentially an automatic decision-making process.

Current machine learning techniques have been widely used in statistical learning (Sutton and Barto 2018), pattern recognition (Neal 2007), data mining (Mitchell 1999), computer vision (Krizhevsky et al. 2012), and natural language processing (Collobert and Weston 2008). In the field of cyberspace security, researchers have also used machine learning for scenarios such as malicious code detection (Huang and Stokes 2016; Liu et al. 2019a), intrusion detection (Debar et al. 1992; Javaid et al. 2016), spam and phishing classification (Abu-Nimeh et al. 2007; Fette et al. 2007), and log analysis (Lane and Brodley 1997; Titonis et al. 2017).

Table 1. The reason why machine learning techniques can be introduced to the fuzzing test process

| Machine Learning | Fuzzing |
| --- | --- |
| Can solve classification problems | There are several classification problems in the fuzzing process, such as whether seed files and test cases are valid, the exploitability of crash, which mutation operator to choose. |
| Long training time and fast decision time | Techniques that consume less performance at run-time are needed so that more test samples can be run in less time. |
| Deep learning can automatically learn grammar semantics | Test input needs to be bypass format check in the program runtime. |
| Need many samples | Fuzzing produces a large number of test samples and crash samples. |
| Input is vector | Natural language processing technology can support, such as one-hot and word2vec. |
| Supervised learning requires label data | Increased code coverage or inclusion of vulnerability paths can be used as labels. |

The application of machine learning technology to fuzzing testing has also attracted the attention of security researchers, and its essence is to treat vulnerability detection as a problem with a program or sample classification (Rajpal et al. 2017; Cheng et al. 2019; Wang et al. 2019). Researchers use existing machine learning techniques to help fuzzer extract experience and knowledge from existing large amounts of vulnerability-related data, and then the new sample is classified and predicted based on the model generated by the training. According to the characteristics of machine learning technology and fuzzing technology, as well as the development of artificial intelligence technology, Table 1 summarizes some of the factors that machine learning technology and fuzzing technology can be combined. Although machine learning technology is used in many industries, there are still many conditions and restrictions on the use of it. As shown in Table 1, the prior condition that machine learning is used to fuzzing has existed, and the new solutions for the problems existing in fuzzing are provided by using machine learning. What problems can be solved by using machine learning and how the effect of the solution in the fuzzing test is discussed in the following sections.

## 4. Applying Machine Learning Techniques for Different Fuzzing Steps

The workflow of fuzzing is shown in Fig. 1. The steps of fuzzing can be classified as follows according to the problems solved by using machine learning:

- Seed file generation
- Testcase generation
- Testcase filter
- Mutation operator selection
- Fitness function
- Exploitability analysis

Table 2 lists the number of corresponding research articles in each category. The total number of references is 29. Testcase generation is the most frequent step in fuzzing process combined with machine learning technology, and the number of research literature is 11. The reason may be that it is easiest to collect a large number of samples and labels for machine learning from the test sample generation step.

Table 2. The distribution of research literatures based on Machine learning for different steps of fuzzing testing

| Step | Number |
| --- | --- |
| Seed file generation | 5 |
| Testcase generation | 11 |
| Testcase filter | 3 |
| Mutation operator selection | 6 |
| Fitness function | 1 |
| Exploitability analysis | 3 |

### 4.1. Seed file generation

The seed file is mutated into fuzzing input sample through various mutation operations. The quality of the input seed file is an essential factor influencing the testing effect. The current seed selection strategy has shortcomings, such as it requires more time to acquire the seed set and the execution effect of selected seeds is almost the same as that of randomly selected seeds (Rebert et al. 2014). The common features of the seed files which lead to higher code coverage, more crashes, and more unique execution paths in the traditional fuzzing test can be learned by using machine learning techniques, and finally generate more seed files with this feature through a generation-based or mutation-based approach.

The data-driven seed generation method implemented by Skyfire (Wang et al. 2017) uses PCFG (Probabilistic context-sensitive grammar, which contains semantic rules and grammatical features) to extract semantic information automatically. These semantic information and grammar rules are used to seed generation. The generated seed file can be guaranteed to pass syntax parsing and semantic checking by using this method. Eventually, Skyfire could execute a deeper path to the target program, thereby more effectively discover deep vulnerabilities.

Fast fuzzing (Nichols et al. 2017) explores the use of deep neural models to enhance the effectiveness of random mutation testing. The method learns features from AFL (Zalewski M 2016) generated samples, and generates seed files that increase the execution path through confrontation training of the Generative Adversarial Networks (GAN).

SmartSeed (Lv et al. 2018) reads and converts the input file into a uniform type of matrix in binary form, and then automatically learns features from the collected datasets that triggering a unique crash or unique path by using WGAN and MLP. Seed files that are easier to cause crashes and unique paths can be generated by the trained model.

Cheng et al. (Cheng et al. 2019) used RNN and seq2seq to find the correlation between the PDF file and the target program execution path. Then, this correlation was used to generate new seed files that are more likely to explore new paths in the target program.

NeuFuzz (Wang et al. 2019) learn the known vulnerability programs and hidden vulnerability patterns in the sample through LSTM to discover the execution path that may contain the vulnerability. Then, NeuFuzz preferentially executes the seed files that can cover the path containing the vulnerability, and assigns more mutation energy to these seed files based on the prediction results.

**4.2. Testcase generation**

A testcase can be generated by performing mutation on the seed file, be constructed based on known input file formats. As the final input, the content of a testcase will directly affect whether or not a bug is triggered. Therefore, constructing a testcase with high code coverage or vulnerability-oriented can effectively improve the efficiency of vulnerability detection in fuzzer.

Samplefuzz (Godefroid et al. 2017) is the first attempt to automatically generate input syntax from sample input using neural networks-based statistical learning techniques. A generation model of automatically PDF object learning based on the seq2seq recursive neural network was proposed and evaluated in this work. The model can generate not only a large number of new useful objects but also improve the coverage.

Fan et al. (Fan and Chang 2017) proposed a method for automatically generating black-box fuzzing testcases for proprietary network protocols. The method uses the seq2seq to learn the generated input model of a proprietary network protocol by processing its traffic, and new messages are generated by using the learning model.

GANFuzz (Hu et al. 2018) learn the protocol syntax by training the generated model in the Generative adversarial network to estimate the underlying distribution function of the industrial network protocol message. A well-formed testcase can be generated based on this generation model.

DeepSmith (Cummins et al. 2018) takes the generation of random programs as a language modeling problem. It learns the syntax, semantics, and conventional structures and patterns of the programming language on the code corpus by using the LSTM model. DeepSmith generated grammatically formatted test samples based on the generation method for fuzzing the compiler.

Sablotny et al. (Sablotny et al. 2018) constructed the model of the stacked RNN to generate HTML tags and new testcases for fuzzing the browser's rendering engine. The main idea is to generate HTML tags based on the probability distribution of learning character sequences from a large number of HTML tags.

IUST DeepFuzz (Nasrabadi et al. 2018) learns the structure of complex input files by using a deep recurrent neural network (RNN)-based neural language model (NLM). IUST DeepFuzz first deletes the non-text part of the input file and replaces it with a token. The token is replaced with a variation of the deleted portion to generate the new testcase after the end of the training.

NEUZZ (She et al. 2019) further proposes a gradient-guided search strategy that calculates and uses the smooth approximation gradient (i.e., NN model) to identify target mutation locations, which could maximize the number of errors detected in the target program. It also demonstrates how to improve the NN model by gradually retraining the model on the wrongly predicted program behavior.

Paduraru et al. (Paduraru and Melemciuc 2018) clustered the corpus with different file formats. By treating the corpus of the input file as a series of characters, the generation model of each cluster is learned by seq2seq. This method can generate new test samples that can perform more branches based on the trained model.

Li et al. (Li et al. 2019b) proposed a method for generating fuzzing test data on industrial control protocols based on WGAN. This method can learn the structure and distribution of real-world data frames and generate similar data frames without knowing the detailed protocol specifications.

DeepFuzz (Liu et al. 2019b) learns the correct C program grammar from the original GCC test suite through the seq2seq model. This model continuously generates grammatically correct C programs based on the learned grammar. Then, the strategies of insert, replace, and remove is used for testcases generation to fuzz the compiler.

V-Fuzz (Li et al. 2019a) constructs a graph-embedded network to train a vulnerability prediction model. The fuzzer is guided to generate samples that tend to reach the area of potential vulnerabilities based on the trained model. The attribute control flow graph of the vulnerability function and the security function is extracted for learning on the function level. The probability of the evaluation is used as metrics for guiding the test sample preference in the prediction stage.

**4.3. Testcase filter**

During the fuzzing test, the PUT needs to execute a large number of samples, and it is time-consuming and inefficient to execute all the samples with uneven quality. The purpose of the testcase filter is to select the test input that is more likely to trigger new paths or vulnerabilities from a large number of samples. Input samples can be analyzed and classified to determine which samples should be further executed to find security vulnerabilities by using machine learning techniques.

Gong et al. (Gong et al. 2017) trained a deep learning model based on samples generated by AFL that caused the program state to be changed as well as the program state to be unchanged. Whether the sample generated by the new round of AFL could change the program state can be predicted by the trained model. Therefore, AFL cannot execute samples that cannot generate new states, which can improve the efficiency of fuzzer.

Augmented-AFL (Rajpal et al. 2017) implements several neural network architectures to learn the capability to predict expected code coverage for a given set of input modifications. In the fuzzing, the learned function is used to predict the heat map of the complete input file, corresponding to the probability of mutations in the location of each file that results in new code coverage. The coverage map is then used to determine the priority of the mutation location.

Siddharth (Karamcheti et al. 2018b) maps program inputs to execution trajectories and sorts the entropy of the execution trajectory distribution. Siddharth is based on the assumption that the higher the uncertainty, the more likely it is to execute a new code path, so the input with the maximal (most uncertain) entropy is selected to perform the next input.

### 4.4. Mutation operator selection

The concept of mutation operators in fuzzing is derived from the biological genetic algorithm. The mutation operator in fuzzing includes operations such as add, modify, and remove. Different mutations at different locations can have different effects. The selection strategy of the mutation operator is to achieve the goals of improving the fuzzing efficiency, such as increasing code coverage or including fragile paths.

Becker et al. (Becker et al. 2010) specified a finite state machine and decomposed different message types to analyze the neighbor discovery protocol. The main idea of the work is to use the reinforcement learning model based on three different reward functions of tracking, debugging, and monitoring networks for fuzzing to choose the best fuzzing test strategy.

LEFT (Fang and Yan 2018) constructs a model based on reinforcement learning to fuzzing LTE functions in Android mobile phones. The model mainly includes three kinds of fuzzing methods: emulation-instrumented black-box fuzzing, threat-model-aware fuzzing, and RL-guided fuzzing.

Inspired by feedback-driven random testing and reinforcement learning, Böttinger et al. (Böttinger et al. 2018) proposed the first fuzzing method using reinforcement learning to maximize code coverage and less processing time. The model can learn the running characteristics of mutations when high returns are achieved.

FuzzerGym (Drozd and Wagner 2018) uses LLVM Santizers' productive and efficient program monitor to obtain status information. This information is used to optimize mutation operator selection using reinforcement learning (RL). The advantages of reinforcement learning and fuzzing are combined to achieve more in-depth coverage across multiple benchmarks by integrating OpenAI Gym with libFuzzer, and realize the learning of the mutation-selection strategy directly from input data.

Karamcheti et al. (Karamcheti et al. 2018a) proposed a Thompson Sampling optimization method based on robbers, which can adaptively adjust the mutator distribution in the process of fuzzing a single program. It is determined which mutation operator should be selected by learning the impact of each mutation operator on code coverage. Finally, the selected mutation operator is used to mutate the test input in the next iteration.

FUZZBOOST (Liu et al. 2019) uses code coverage information collected from runtime traces as a reward and optimizes this reward using the Deep Q-learning algorithm. By doing this, the fuzzing agent learns how to select mutation operators for mutating the seed program to improve the coverage of fuzzing tests.

### 4.5. Fitness function

In genetic algorithms, a fitness function is a particular type of objective function that is used to summarize, as a single figure of merit, how to close a given design solution to achieve the set aims (Wikipedia 2019a). Fitness function refers to the evaluation method used to distinguish the satisfactory and unsatisfactory standards of testcase in the state-of-the-art fuzzer based on genetic algorithm. Common fitness functions include code coverage, potential vulnerability position.

Xiao et al. (Sun et al. 2018) proposed a new genetic programming-based fitness function, which is different from the current mainstream code coverage-based method. It combines Markov-chain and PCFG model to learn commonness from a corpus of normal scripts developed by programmers, and uses the learned information to compute the script uncommonness by measuring the deviation of the script to a common script. Scripts with larger deviation may be more likely to trigger errors in the interpreters. This deviation is used to compute the fitness of the GP-based language fuzzing script.

### 4.6. Exploitability analysis

Vulnerability exploitability refers to the possibility of the vulnerability being attacked and exploited by the attacker. It is an inherent property of vulnerability. In fuzzing, there are a lot of crashes and error messages, but a few of them are vulnerabilities. How to find real vulnerabilities from these crashes is a challenge. Commonly used vulnerability analysis methods are static analysis and dynamic analysis. Adopt the technology of tools such as !exploitable(Team MSECMSS 2013), CERT tools.

ExploitMeter (Yan et al. 2017) uses the Bayesian machine learning algorithm to make initial judgments on the static features extracted from the software. The initial judgments and the exploitability judgments in the fuzzing process are combined to update the final Exploitability results.

Exniffer (Tripathi et al. 2018) suggests using machine learning to determine more general rules for crash exploitability prediction automatically. The method uses support vector machines (SVM) to learn the features extracted from core dump files (generated during crashes) and information from the most recent processor hardware debugging extensions.

Zhan et al. (Zhang and Thing 2019) generated compact fingerprints for dynamic execution tracking of each crash input based on n-gram analysis and feature hashing. The fingerprint is then fed to an online classifier to build a distinguishing model. Incremental learning enabled by online classifiers allows models to scale well even for large numbers of crashes, while being easy to update for new crashes.

## 5. Analysis of Machine Learning Based Fuzzing Model

The work that systematic comparison of the performance of various algorithms is lesser in the present work of choice of machine learning algorithms for fuzzing. This section systematically summarizes the knowledge of the machine learning model used in the fuzzing test. It summarizes the following five aspects:

- Selection of machine learning algorithm
- Pre-processing methods
- Datasets
- Evaluation metrics
- Hyperparameters setting

### 5.1. Selection of machine learning algorithm

The fuzzing test utilizes the classification capability of machine learning to train hidden vulnerability detection models. The training data is from a large number of known sample sets and program execution feedback information, which can effectively improve the efficiency of vulnerability detection. However, different machine learning algorithms are applied to different scenarios. Even in the same scenario, choosing different algorithms can lead to significant differences in results (Zou et al. 2018). The input data of the fuzzer can be hexadecimal text, source code, binary string, network packet, and other forms. The PUT also contains complex syntax, semantics, and logical structure. It is a tough problem to judge which machine learning algorithm effective for the complex environment of the fuzzing test.

Table 3 lists the machine learning algorithms and their distributions used in the fuzzing. The first column shows the name of the machine learning algorithms, and the second column counts the number of times the corresponding algorithm was used (appears and implemented in the literature). The third column indicates the category to which the algorithm belongs, including three categories: traditional machine learning, deep learning, and reinforcement learning. Each traditional machine learning algorithm is used only once. The reason for this phenomenon may be that traditional machine learning techniques require manual extraction of features. However, both the input sample format and the target program contain complex syntactic and semantic structures, and there are no valid vulnerability models or vulnerability features.

Deep learning relies on its representation learning to have the capability to automatically extract features for a wide range of applications in fuzzing testing. The two most used algorithms are LSTM and seq2seq, which are used 9 times and 6 times, respectively. The reason that LSTM is used the most is its excelling at processing sequential data: the program execution path is very similar to the statement in natural language, and whether a piece of code contains a vulnerability depends on the context. On the other hand, LSTM has a memory function suitable for handling long dependencies because the code associated with the vulnerability may be located at a relatively long distance in the path(Wang et al. 2019). The length of input and output sequences of the seq2seq model is variable, which can effectively use the input of fuzzing as text data to learn local or global syntax information. New neural networks, such as Generative adversarial network (Goodfellow et al. 2014) and Graph Convolutional Network (Kipf and Welling 2016), is also used in the fuzzing test.

Reinforcement learning is used for the selection of mutation operators in the fuzzing test because reinforcement learning needs to choose different actions in different environments, which is similar to the selection of mutation operator. However, reinforcement learning itself has limitations, such as long training time, weak convergence, and local optimization, which leads to its less used in the fuzzing test (Mnih et al. 2013).

Table 3. Machine Learning Algorithm Distribution for Fuzzing

| Algorithm | Number | Category |
|---|---|---|
| LR (*Logistic Regression*) | 1 | |
| NB (*Naive Bayes*) | 1 | |
| SVM (*Support Vector Machines*) | 1 | |
| RF (*Random Forest*) | 1 | |
| DT (*Decision Tree*) | 1 | Traditional Machine learning |
| PCFG (*Probabilistic Context-Free Grammar*) | 1 | |
| PCSG (*Probabilistic Context-Sensitive Grammar*) | 1 | |
| PA (*Passive-Aggressive*) | 1 | |
| Thompson Sampling | 1 | |
| RNN (*Recurrent Neural Network*) | 3 | |
| CNN (*Convolutional Neural Network*) | 2 | |
| **LSTM (*Long Short Term Memory*)** | **9** | |
| GRU (*Gate Recurrent Unit,*) | 1 | |
| BLSTM (*Bidirectional Long Short-Term memory*) | 2 | Deep learning |
| **Seq2seq (*Sequence-to-sequence*)** | **6** | |
| MLP (*Multilayer Perceptron*) | 2 | |
| GCN (*Graph Convolutional Network*) | 1 | |
| GAN (*Generative Adversarial Networks*) | 2 | |
| WGAN (*Wasserstein Generative Adversarial Networks*) | 2 | |
| Q-Learning | 1 | |
| SARSA (*State–action–reward–state–action*) | 1 | Reinforcement learning |
| Deep Q-Learning | 2 | |
| Deep Double Q-Learning | 1 | |

## 5.2. Pre-processing method

Due to the different types of PUT, the input format is quite different, such as text, pictures, video, network data packets, and program code. This data needs to be converted into an input that can be used for machine learning. Table 4 summarizes the data preprocessing methods commonly used for fuzzing tests.

Table 4. Pre-processing Method for Machine Learning Technology in Fuzzing

| Pre-processing Method | Description |
|---|---|
| Program analysis | The extracted information is transformed into vectors through static or dynamic analysis |
| Natural language processing | Direct use of n-gram, count statistics, Word2vec, heat map, and other ways to convert the input into a vector. |
| Others | A combination of program analysis and natural language processing, custom, take the entire file or component element as input |

In fuzzing, pre-processing methods are divided into three categories: program analysis, natural language processing, and others. Program analysis refers to extracting program features or runtime information, such as stacks, registers, assembly instructions, jumps, program control flow graphs, abstract syntax trees, and program execution paths, by techniques using static or dynamic analysis (Wang et al. 2017; Tripathi et al. 2018; Li et al. 2019a). Natural language processing refers to the methods directly letting input as text, using sophisticated text processing techniques to extract hidden features in the input data, such as n-gram (Damashek 1995), count statistics, Word2vec (Goldberg and Levy 2014), heat map(Wilkinson and Friendly 2009) and other methods(Fan and Chang 2017; Yan et al. 2017; Karamcheti et al. 2018b). Others include combining program analysis with natural language processing techniques (Wang et al. 2019; Zhang and Thing 2019) or converting entire documents or pdf objects into vectors (Rajpal et al. 2017; Lv et al. 2018), as well as custom methods. As defined in the literature (Gong et al. 2017), the binary sequence of testcase is represented by 32-bits, fuzzing technique is represented by 4-bit, mutation bits is represented by 10-bit, and the mutation value is represented by 32-bits, whether the new test case is represented by 1-bit. Finally, each piece of data can be combined into a 79-bit binary sequence, with the first 78-bit as input and the last 1-bit as a label.

### 5.3. Datasets

The performance of machine learning is primarily influenced by the training data. Especially, deep learning can easily lead to over-fitting when the amount of data is insufficient. In the present work, the datasets used for machine learning algorithm based fuzzing test are the following sources:

- Web-crawler
- Fuzzing generation
- Self-build
- Public dataset

Web crawlers (Menczer et al. 2001) are commonly used methods for collecting data, especially for widely used file formats such as DOC, PDF, SWF, and XML. Conventional crawling methods can be downloaded according to specific file extension filter conditions, specific magic bytes and other signature methods (Godefroid et al. 2017; Wang et al. 2017; Cheng et al. 2019).

The fuzzing generation is to executing a similar fuzzer such as AFL and collects the generated samples and their tag data (coverage, code execution path, etc.) for a period of time. This method can generate datasets in various formats, and the number of samples can be satisfied (Gong et al. 2017; Rajpal et al. 2017; Lv et al. 2018; She et al. 2018).

The self-build approach is similar to the fuzzing generation, but it uses other means, such as a self-built communication environment to grab traffic packets as datasets (Fan and Chang 2017; Hu et al. 2018; Li et al. 2019b).

The public datasets used and their corresponding categorizations are as follows:

- **Learning from "Big Code" datasets**(Github 2017): At present, as an open-source project, making the above contains a large amount of public data associated with the code, such as Python ASTs(This dataset includes 100'000 + 50'000 python files as parsed abstract syntax trees along with the code of the parser), JavaScript ASTs(This dataset includes 150,000 JavaScript files. The data is available as JavaScript and as parsed abstract syntax trees).

- **The NIST SARD project datasets** (NIST 2006): It is the ground truth of examining the problem code as a software assurance tool, which contains test cases of over 100,000 different programming languages, covering dozens of different categories of weaknesses, such as those in the Common Weakness Enumeration (CWE). These test cases contain files for all phases of the software lifecycle, such as design, source code, and binaries, which also contain patches for weaknesses.

- **GCC test suite datasets** (GCC 2019): The GNU Compiler Collection includes front-ends for C, C++, Objective-C, Fortran, Java, Ada, and Go languages, as well as libraries for these languages (such as libstdc++, libgcj.).

- **DARPA Cyber Grand Challenge datasets** (DARPA CGC 2016): The DARPA Network Challenge Binaries is a set of 200 binary programs with extensive functionalities released by DARPA. These programs are part of an open challenge for creating tools that automatically modify, validate, and fix errors. Common to all of these binaries is that each binary contains one or more bugs that are generated by humans when they programming, which is documented by the developers.

- **LAVA-M datasets** (Dolan-Gavitt et al. 2016): Consisting of four programs from the GNU Coreutils suite: uniq, base64, md5sum, and who, which are injected with 28, 44, 57, and 2265 errors with unique IDs, respectively, with some unlabeled errors. These errors are located deep in the program and are only triggered when an offset in the program input buffer matches a 4-byte "magic" (random) value. This data set has become popular in recent years for benchmarking complex white box fuzzers, symbolic execution tools, and some gray box fuzzers.

• **VDiscovery datasets** (Grieco et al. 2016): The dataset is released by VDiscovery which is a vulnerability discovery tool, and contains a total of 402 unique samples. These samples consist of 138,308 sequences of system calls for 1,039 Debian programs.

Table 5. Evaluation Metrics and Details for Machine Learning Based Fuzzing Model

| Performance metrics | Description | count |
|---|---|---|
| Accuracy | It is the proportion of the total number of correct predictions amongst the total number of correct as well as incorrect predictions. | 5 |
| Precision | It is the proportion of correctly classified fault-prone classes amongst the total number of classified fault prone classes. | 5 |
| Recall (true positive rate (TPR)) | It is the proportion of correctly predicted fault prone classes amongst all actual fault-prone classes. | 4 |
| TNR (true negative rate) | It is the proportion of correctly predicted non-fault prone classes amongst all actual non-fault prone classes. | 1 |
| FPR (false positive rate) | It is the proportion of all non-fault prone classes which are incorrectly predicted as fault-prone. | 3 |
| FNR (false negative rate) | It is the proportion of faulty classes that are classified as non-fault prone. | 2 |
| ROC | ROC (receiver operating characteristic curve) is plotted with TPR values on the y-axis and the FPR values on the x-axis. | 2 |
| F-measure | It is the harmonic mean of precision and sensitivity. | 3 |
| Loss | It represents some function of the difference between estimated and true values for an instance of data | 4 |
| Models perplexity | The perplexity shows the difference between the predicted sequence and test set sequence. | 1 |

Table 6. Evaluation Metrics and Details for Fuzzing Model Based on Machine Learning

| Performance metrics | Description | count |
|---|---|---|
| Coverage | It includes instruction coverage, basic block coverage, line coverage, branch coverage, edge coverage, function coverage, and relative coverage | 15 |
| Unique code path | It represents the number of code paths that are executed or triggered during testing. | 4 |
| Unique crash or bug | It represents the number of unique crashes and bugs found during testing. | 13 |
| Pass rate | It represents the proportion of samples generated that can be verified by the syntax of PUT | 3 |
| Efficiency | It represents the measures of time consumption during testing, such as training time of machine learning model, seed generation speed, sample generation speed, execution time, and so on. | 12 |

## 5.4. Evaluation metrics

The performance evaluation of the fuzzing methods based on machine learning technology can be divided into two aspects: the evaluation of the performance of the machine learning model and the evaluation of the vulnerability detection capability.

The evaluation of the machine learning model is based on the classification metrics. Table 5 summarizes the metrics and detailed information used to evaluate the machine learning model in fuzzing, and the number of times these metrics were used was also counted. According to the statistics in Table 5, the most commonly used performance metrics are Accuracy and Precision, which are closely followed by Recall, Loss, FPR, and F-measure. FPR, Models perplexity is the least used.

The evaluation of vulnerability detection capability of the fuzzing method based on machine learning is the same as that of traditional fuzzing methods. Table 6 summarizes the metrics and details of the vulnerability detection performance that have been used to evaluate the fuzzing method based on machine learning technology.

Both the traditional fuzzing method and the machine-learning-based fuzzing method are designed to find vulnerabilities. So coverage, unique crashes, and bugs are valid metrics for evaluating the performance of the fuzzing model. However, the fuzzing method based on machine learning has model training, feature extraction, and other steps, so efficiency is also used many literatures.

### 5.5. Hyperparameters setting

In the implementation of the machine learning model, the value of hyperparameters is not obtained through training but requires artificial settings before training. In general, it is necessary to optimize the hyperparameters and select an optimal set of hyperparameters to improve the performance and effectiveness of learning. Table 7 analyzes and compares the works of literature, and summarizes the values that have been selected for some crucial hyperparameters in machine learning.

Table 7. Analysis of Hyperparameters Setting of Machine Learning based Fuzzing Model

| Hyperparameters | Selected value | summary |
| --- | --- | --- |
| Number of layers | 1-8 | Select a maximum of 4 layers and 2 hidden layers. |
| Number of nodes in each layer | 32, 64, 100, 128, 256, 512, 1024, 4096, 8192 | Among them, 128 and 256 are more. |
| Epochs | 5, 10, 15, 20, 30, 40, 50 | The maximum epoch's option is 50, but 40 works best. |
| Activation function | Sigmoid, elu, softplus, softsign, ReLU, tanh | Sigmoid, ReLU, and tanh are most commonly used. |
| Learning rate | 0.0001, 0.0005, 0.001, 0.002, 0.02, 0.1 | 0.001 was used the most |

The hyperparameters in the deep learning algorithm are mainly selected to complete the comparison, including the number of layers, number of nodes in each layer, epochs, activation function, and learning rate. The different number of layers and number of nodes in each layer will affect the accuracy and complexity of the whole neural network. Over-fitting will also occur when the number of nodes in each layer is large. In the fuzzing scenario, the maximum number of layers is 4, and the number of nodes are 128 and 256. As the number of epochs increases, the weight updating iterations of the neural network will increase, and the loss function curve will enter the optimized fitting state from the initial unfitting state to the over-fitting state. Usually, the maximum number of epochs selected is 50, but the best effect can be reached at 40. The choice of the activation function could improve the ability of the neural network to model expression, and solve the problem that cannot be solved by the linear model. However, the advantages and disadvantages of different activation functions are different, such as sigmoid input range between [0, 1], but there are "sigmoid saturate and kill gradients" and not "zero-centered" problems (LeCun et al. 2012). Tanh (Fan 2000) solves the sigmoid output of not "zero-centered", but other problems still exist. In fuzzing, Sigmoid, ReLU, and Tanh are most commonly used activation functions. The learning rate controls the learning progress of the model and also affects the speed of the model's convergence to the local minimum value. A higher learning rate will easily lead to an explosion and shock of loss value, while a lower learning rate will lead to slow over-fitting and convergence speed. Many values of learning rate are selected in fuzzing, and 0.001 is used more.

## 6. Performance Evaluation of Fuzzing Model Based on Machine Learning

The performance evaluation of machine learning-based fuzzing model is divided into two parts: on the one hand, evaluating the performance of the machine learning model used in fuzzing to discuss whether machine learning technology is with reasonable classification ability in the scenario of fuzzing test. On the other hand, evaluating the vulnerability detection ability of the fuzzing model to discuss whether the vulnerability detection capability is improved by using machine learning technology. Considering that the detailed parameter settings of the experiments are not explicitly given in many works of literature, and the experimental codes are not open-sourced, the data used in this section are derived from the experimental data in the corresponding literature.

### 6.1. Performance evaluation of machine learning model for fuzzing

In this section, we summarize the results of the machine learning based fuzzing model. The results in Section V show that Accuracy, Precision, Recall, and Loss are the most frequently used performance measures in the selected research literatures. We provide the results of these four performance measures, where Accuracy, Precision, Recall values are selected as the maximum value of the model determined in each literature, the value of Loss is selected as the minimum value of the model determined in the corresponding literature.   The results are shown in Fig. 2, 3, 4, and 5.

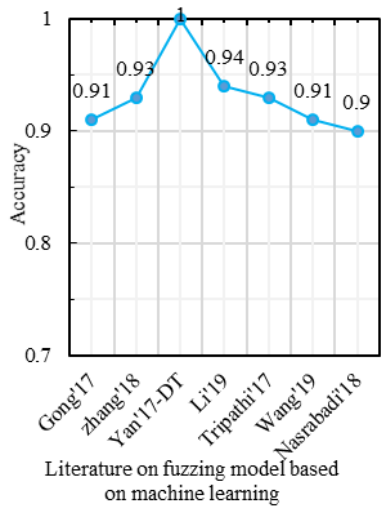
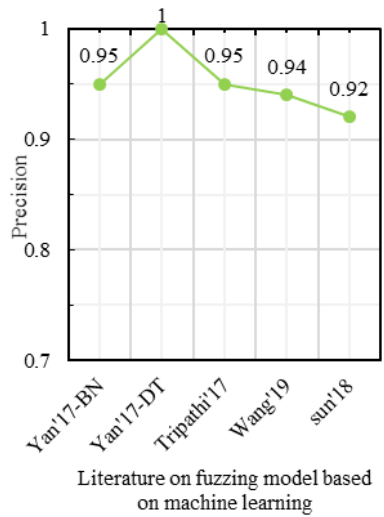

**Fig. 2. Comparison of Accuracy between different models**  **Fig. 3. Comparison of Precision between different models**

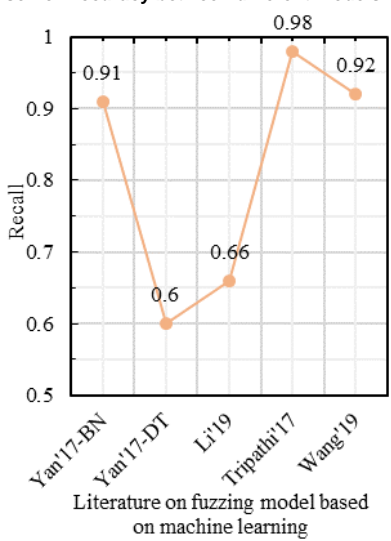
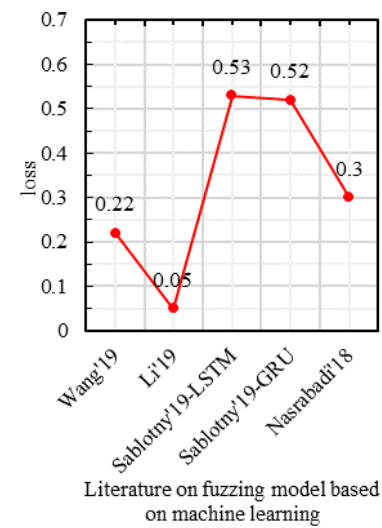

**Fig. 4. Comparison of Recall between different models**  **Fig. 5. Comparison of loss between different models**

The higher value of the three types of performance metrics, including accuracy, precision, and recall, indicates a more accurate prediction. In terms of accuracy, the value of the machine learning model in the listed literature is above 0.9. Precision results in Fig. 3 also show that the machine learning model in the listed literature has a precision value above 0.92. The statistical results of the performance metrics recall in Fig. 4 show that the lowest recall value in the listed literature is 0.6, and the highest recall value is 0.98. Especially, the values of the accuracy and precision are 1 in the Exploitmeter proposed by Yan'17-DT (Yan et al. 2017), which is the highest among the experimental results of multiple types of features. However, all of the experimental results of Exploitmeter average accuracy is 0.9, the precision of the average of around 0.4, and average recall is 0.2. The accuracy of V-fuzz proposed by Li '19 (Li et al. 2019a) can reach more than 0.9 after the experiment is stable, but its recall value is also low, only 60%. The main reason for the occurrence of the values of these metrics in these two models is that the features selected in the paper are not highly correlated with vulnerabilities, so they cannot be effectively used as the prediction of exploitable samples and vulnerability samples. However, the lower the value of the performance metrics Loss, the higher the robustness of the model. The statistical recall value in Fig. 5 is at least 0.05, and the highest is 0.53. The values of these performance metrics indicate that the machine learning model for fuzzing has reasonable predictive power.

### 6.2. Performance evaluation of vulnerability detection capability

In fuzzing test, there are challenges such as how to mutate and generate input seed files, how to increase coverage rate, and how to pass the validation effectively. Whether the application of machine learning technology can effectively alleviate some bottlenecks is the critical problems is studied in this paper. This chapter summarizes the experimental results of the fuzzing tools based on machine learning and evaluates them from the aspects of coverage, unique code path, unique crash or bug, Pass rate, and efficiency.

### 6.2.1 Coverage

Coverage is the most frequently used metrics in the fuzzing test to evaluate the vulnerability detection performance. Table 8 summarizes the experimental results of all the machine learning-based fuzzing models mentioned in this paper. On the whole, the application of machine learning technology in fuzzing improves code coverage. Especially, NEUZZ has a fourfold increase in coverage compared to baseline AFL (She et al. 2018). However, there are some models, such as V-Fuzz (Li et al. 2019a), Cheng'19 (Cheng et al. 2019), SampleFuzz (Godefroid et al. 2017), and Augmented-AFL (Rajpal et al. 2017), which did not improve much comparing with the baseline. The reason can be summarized into three points: (1) not oriented for increasing coverage, but with a vulnerability-oriented driver process such as a vulnerable path; (2) the tension between the coverage and the pass rate, that is if the format of the generated sample is suitable, the coverage rate is low, and conversely, the coverage rate is high; (3) model query and execution trade-offs, the larger file will consume more time when the prediction process is performed before the fuzzing test. So in order to execute more samples in a short time, the size of selected sample files should be small, so that the code executed is limited.

Table 8 Results of Coverage Improvement Fuzzing Model Based on Machine Learning

| Coverage category | Fuzzing model | comparison | result |
|---|---|---|---|
| basic block coverage | Neural Fuzzing | AFL, Augmented-AFL, SampleFuzz | Increased by 2.26%, 7.73%, and 7.56% respectively. |
| | Augmented-AFL | AFL | Significant improvements in code coverage metrics for readelf and readpng programs, a maximum increased by 1.26%, but in the other two programs, performance is not much different. |
| | V-Fuzz | VUzzer | Little difference. |
| | FUZZBOOST | GCC test suite baseline | Increased by 37.14%. |
| | Cheng'19 | AFL | Increased by 2.18% |
| line coverage, | FuzzerGym | libFuzzer | Up to about 2.5 times more line coverage than libfuzzer. |
| | Skyfire | Crawl+AFL | Increased by 20% |
| | DEEPFUZZ | Csmith | Increased by 6.69% |
| relative coverage | Karamcheti'18' | FidgetyAFL, Batched FidgetyAFL | Higher code coverage is achieved, and as the number of executions increases, the gap grows. When executed up to 50,000 times, the relative coverage can be increased by about 10%. |
| | Thompson Sampling | AFL, FidgetyAFL, FaireFuzz | Increased by 10%. |
| function coverage | Skyfire | Crawl+AFL | Increased by 15%. |
| | DEEPFUZZ | Csmith | Increased by 2.26%. |
| instruction coverage | SampleFuzz | Normal sample execution | Increase by about 1%. |
| | Cheng'19 | SampleFuzz | The new corpus covers 0.11% more instructions. |
| branch coverage | NEUZZ | AFL, AFLFast, VUzzer, KleeFL, AFL-laf-intel, RNN | It is 4 times higher than the AFL and 2.5 times higher than the AFLFast with the second-highest performance. Edge coverage of 3.7x to 8.4x is achieved in different projects compared to RNN-based fuzzes. |
| | DEEPFUZZ | Csmith | Increased by 2.36%. |

### 6.2.2 Unique code path

In terms of finding unique code paths. Augmented-AFL has significant gains compared to AFL in readpdf and readelf, which is up to 2 times, but no significant improvements in mupdf and libxml. Faster Fuzzing does not cause deep path exploration when the synthetic seed file is provided as input to the program compared to AFL. The model proposed by Cheng '19 et al. (Fan and Chang 2018) explores an average of 24.30% more execution paths than AFL on mupdf programs. SmartSeed(Lv et al. 2018) finds the most paths in 8 programs, ranked second 1 program, and ranked third in 3 programs among all the 12 target applications. SmartSeed + AFL find 30.7% unique paths of more than the best seed strategy available.

### 6.2.3 Unique crash or bug

Table 9 summarizes the results of crash and bug experiment in different works of literature. The fuzzing tools that introduce machine learning can find more crashes and bugs than the traditional fuzzing tool in the actual program and exposes multiple CVEs on the whole. However, there are also some tools, such as Neural fuzzing, that fail to find any crash during the experiment.

Table 9 Machine Learning-Based Fuzzing Compared to Baseline Results in Unique Crash and Bug Discovery

| Fuzzing model | Comparison | Result |
| --- | --- | --- |
| SampleFuzz | -- | One bug was discovered during the five-day experiment. |
| Augmented-AFL | AFL | More than 20 and 110 unique crashes were found on readelf and libxml programs, respectively. |
| NeuFuzz | PTfuzz, QAFL | On the LAVA-M dataset, 19 crashes were discovered, and more crashes were found than the 3 found by PTFuzz and 0 found in QAFL. In the real program, 42 vulnerabilities were discovered, 14 of which were disclosed by the predecessors, 21 of which were confirmed by CVE. |
| Skyfire+AFL | Crawl | Skyfire+AFL found 19 previously undiscovered memory crashes, 21 new denials of service bugs, more than Skyfire or Crawl+AFL. 11 of these vulnerabilities were identified as CVE. |
| DeepSmith | CLSmith | 29 different compiler assertions are fired. DeepSmith also triggered 3 unreachable compile-time crashes that CLSmith did not trigger. |
| Cheng'19 | AFL | 67 crashes were triggered, including 2 CVE vulnerabilities, while AFL's original seed corpus triggered only 32 crashes, with no CVE vulnerabilities. |
| NEUZZ | AFL, AFLFast, VUzzer, KleeFL, AFL-laf-intel | 60 bugs were found in 6 different programs, more than AFL (29), AFLFast (27), VUzzer (7), KleeFL (14), and AFL-laf-intel (26). |
| V-Fuzz | Vuzzer, AFL, AFLFast | V-Fuzz found the most unique crash, compared to VUzzer, the average number of unique crashes discovered by V-Fuzz increased by 35.8%; a unique crash was discovered faster than other fuzzers; On LAVA-M, V-Fuzz found more crashes; V-FUZZ found 10 CVEs, 3 of which are new. |
| SmartSeed+AFL | Random, AFLresult, Peachset, Hotset, AFL-cmin + AFL | Unique crash of 124.6% was found over the best seed strategy available, and 16 CVE vulnerabilities were discovered. |
| Neural Fuzzing | -- | No crash found |
| DeepFuzz | Csmith | DeepFuzz found 8 newly confirmed GCC errors. |
| Thompson Sampling | AFL, FidgetyAFL, FaireFuzz | Thompson Sampling found 1336 unique crashes in all 75 test binaries, almost twice the number found by any other fuzzer. |

Table 10 Machine Learning-Based Fuzzing Model and State-of-the-art Fuzzing Statistics on The Number of Vulnerabilities Discover on LAVA-M Datasets

|  | base64 | md5sum | uniq | who |
| --- | --- | --- | --- | --- |
| #Bugs | 44 | 57 | 28 | 2136 |
| AFL | 0 | 0 | 9 | 1 |
| VUzzer | 17 | -- | 27 | 50 |
| FUZZER | 7 | 2 | 7 | 0 |
| SES | 9 | 0 | 0 | 18 |
| Steelix | 43 | 28 | 24 | 194 |
| Angora | 48 | 57 | 29 | 1,541 |
| AFL-laf-intel | 42 | 49 | 24 | 17 |
| InsFuzz | 48 | 38 | 11 | 802 |
| T-fuzz | 43 | 49 | 26 | 63 |
| REDQUEEN | 44 | 57 | 28 | **2134** |
| DigFuzz | 48 | 59 | 28 | 167 |
| NEUZZ | **48** | **60** | **29** | 1,582 |
| NeuFuzz | 6 | -- | 5 | 8 |
| Thompson Sampling | 31 | 1 | 0 | 106 |
| V-Fuzz | 27 | -- | 28 | 62 |

Table 10 summarizes the number of bugs found by the traditional fuzzing tools and the machine learning-based fuzzing tools based on the LAVA-M dataset. The first line lists the four programs in LAVA-M, the second line lists the number of bugs exposed in each program, and the third line to the last line indicates the number of bugs found by each different fuzzing tools in the four programs of LAVA-M. The last four lines are the experimental results of the fuzzing tools using machine learning methods. From the statistical data, it can be found that the vulnerability discovery capability of the machine learning-based fuzzing tool is not improved compared with the state-of-the-art fuzzing tools, such as REDQUEEN(Aschermann et al. 2019), DigFuzz(Zhao et al. 2019), Angora(Chen and Chen 2018), InsFuzz(Zhang et al. 2019) and T-fuzz(Peng et al. 2018). The only better one is the NEUZZ tool, which can maintain the highest number among three of the four programs. In general, the fuzzing method based on machine learning has not substantially improved the vulnerability detection capability compared with the traditional fuzzing method. However, there is a threat to the validity of this conclusion. Due to the different goals, fewer the machine learning-based fuzzing tools can test the LAVA-M dataset, so it is difficult to obtain and summarize the conclusions.

### 6.2.4 Pass rate

The pass rate represents the percentage of generated samples that can pass the program's syntax check. Comparing to the 34% pass rate of CFG on XML, Skyfire (Wang et al. 2017) has 85% XSL and 63% XML that can pass semantic detection and reach the application execution state due to its consideration of context. The pass rate of SampleFuzz (Godefroid et al. 2017) is between 70% and 97%, showing good learning quality. The pass rate of DeepFuzz (Liu et al. 2019b) increases with the number of iterations. The optimal pass rate of all sampling methods is achieved in 30 iterations of training. The highest pass rate is 82.63%, and the training is stable after 80%. Sample files generated by deep learning and automatic learning based on grammatical semantic information have a high sample pass rate. But finding a security vulnerability requires executing a file whose format is corrupted, so it is necessary to weigh the proportion of sample validity.

Table 11. Evolution Results of Efficiency Based on Machine Learning Fuzzing Model

| Efficiency | Fuzzing model | Comparison | Result |
|---|---|---|---|
| Execution time | Gong'17 | AFL | On average, the time overhead increased by about 5%. |
| | NeuFuzz | QAFL, PTFuzz | In terms of execution efficiency (number of test cases executed per second), it is 2.5 times faster than QAFL and about 8% slower than PTFuzz. |
| | Bottinger'18 | random selection | By combining time and coverage as rewards for reinforcement learning, execution time was improved by 11.3% |
| | DeepSmith | CLSmith | DeepSmith averages 4.46 times faster than CLSmith in test case execution efficiency. |
| Generated time | Faster Fuzzing | random selection, LSTM | The seed file generated by the GAN-based method is 14.23% faster than the random method and 60.72% faster than the LSTM. |
| | SmartSeed | random selection, peachset, AFL-cmin, | Random selection is the fastest. SmartSeed takes 12 seconds to generate 100 seed files and 240 seconds to generate 2000 seed files. For example, to generate 500 seed files, it takes peachset and AFL-cmin 2,500 seconds and 874 seconds respectively, and hotset needs >2,000 minutes to select seeds from 500 files in our experiment. |
| | DeepSmith | CLSmith | The generation time of DeepSmith increases linearly with the length of the program, which is 2.45 times faster than CLSmith. |
| | Skyfire | -- | Skyfire generates 18,686, 19,324, and 525,647 XSL, XML, and JavaScript seeds requiring only 20.3 seconds, 20.6 seconds, and 521.2 seconds, respectively. |

### 6.2.5 Efficiency

The test efficiency is to evaluate the time overhead of the fuzzing runtime, including two aspects: execution time and generation time. Execution time refers to the time overhead of executing test samples in the fuzzing process, such as the time it finds a given number of crashes and hangs, the number of test samples executed per second, and the time it takes to execute the same test sample. The generation time refers to the time it takes to generate a seed file or a test sample. Table 11 summarizes the experimental results of the test efficiency of the machine learning-based fuzzing models. In terms of execution time, using machine learning can select high-quality samples in advance and reduce the execution of useless samples. By doing this, fuzzer improves the efficiency of the fuzzing test. However, the execution speed of the Neufuzz (Wang et al. 2019) is about 8% slower than PTfuzz (Zhang et al. 2018) due to the need for path recovery. As for the generation time, since the grammar of the program input is learned by the machine learning method, files can be generated faster than traditional methods such as checking and filtering. In general, the fuzzing model using machine learning has a good improvement in test efficiency compared to the traditional fuzzing model.

## 7. Conclusion and Future Directions

In this paper, we systematically review the works of literature to analyze and assess the performance of machine learning techniques for fuzzing. First, we introduced the concept of fuzzing and the key challenges that currently exist. Second, we analyzed and summarized the reasons why machine learning technology can be used in fuzzing scenarios. Third, we emphatically summarized the use of machine learning techniques for different stages, such as seed file generation, testcase generation, testcase filter, mutation operator selection, fitness function, and exploitability analysis. Fourth, we summarized the characteristics of the primary research from the perspectives of selection on machine learning algorithm, pre-processing methods, datasets, evaluation metrics, and hyperparameters settings. Then the performances of the machine learning based fuzzing models are assessed based on four metrics (accuracy, precision, recall, and loss). The results depict that machine learning has an excellent predictive capability for fuzzing. Finally, the vulnerability discovery capability of the machine learning-based fuzzing tool is analyzed by comparing with the traditional fuzzing tool. By Comparing coverage, unique code path, unique crash or bug, pass rate, and efficiency, the results depict that the introduction of machine learning technology in fuzzing can improve the performance of fuzzing. In some ways, however, machine learning based fuzzing tools has some drawbacks compared to the state-of-the-art fuzzing tools.

Future directions can be carried out from the following three aspects:

### (1) Datasets

In this study, we find that there is no public dataset that can be used as a benchmark in the current fuzzing field. Datasets constructed from web crawlers, fuzzing generation, self-builds are not universal and are less recognized. Some public data sets have been used to contain fewer categories, fewer features, and the data used for training is unbalanced. The quality of the data set will seriously affect the performance of the vulnerability detection model. Therefore, we believe that the introduction of machine learning into fuzzing must establish open data sets that can be used as a test benchmark.

### (2) Feature selection

Machine learning constructs a classification model by learning the characteristics of the data sets. The selection of different features will lead to different classification accuracy and precision of the model. The structure of the program and the execution information are not directly related to the vulnerability in the field of vulnerability discovery, so how to select practical features from the program or sample becomes an essential factor affecting the performance of fuzzing. At present, the natural language processing technology is relatively mature, so we can consider using advanced technologies in the field of natural language processing to extract useful information such as code attributes, semantic and grammatical features of programs for fuzzing.

### (3) Selection of learning algorithms

Different machine learning techniques are suitable for different scenarios, and different network configurations can lead to different results. First of all, the characteristics of different stages of fuzzing, the size of the corresponding data, the advantages and disadvantages of different algorithms should be used as the basis for the algorithm selection. Secondly, graph convolutional networks, fusion neural networks, and interpretable deep learning models can all be tried to integrate with fuzzing, and it is necessary to study more complicated and suitable neural network models to improve the quality of generated samples.

## ACKNOWLEDGMENT

This work was supported by the National Key Research and Development Program of China under Grant 2017YFB0802900.